# Strong interlayer coupling in two-dimensional PbSe with high thermoelectric performance

Z. P. Yin, C. Y. Sheng, R. Hu, S. H. Han, D. D. Fan, G. H. Cao, and H. J. Liu[*]

*Key Laboratory of Artificial Micro- and Nano-Structures of Ministry of Education and School of Physics and Technology, Wuhan University, Wuhan 430072, China*

It was generally assumed that weak van der Waals interactions exist between neighboring layers in the two-dimensional group-IV chalcogenides. Using PbSe as a prototypal example, however, we find additional strong coupling between the Pb-Pb layers, as evidenced by detailed analysis of the differential charge density. The coupling resembles covalent-like bond and exhibits strong harmonicity around the equilibrium distance, which can be fine tuned to obviously reduce the phonon thermal conductivity but slightly change the electronic transport of PbSe. As a consequence, a maximum *ZT* value of 2.5 can be realized at 900 K for the *p*-type system. Our work offers an effective and feasible design strategy to enhance the thermoelectric performance of similar layered structures.

## 1. Introduction

Thermoelectric (TE) materials can directly convert heat into electrical energy without generating other by-products, which have attracted much attention from scientific community for their environmentally friendly advantages [1]. Usually, the performance of a TE material can be described by the dimensionless figure-of-merit $ZT=S^2\sigma T/(\kappa_e+\kappa_l)$ [2-4], where $S$ is the Seebeck coefficient, $\sigma$ is the electrical conductivity, $T$ is the absolute temperature, $\kappa_e$ is the electronic thermal conductivity, and $\kappa_l$ is the lattice thermal conductivity, respectively [5,6]. An ideal TE material must possess high power factor ($PF=S^2\sigma$) and low thermal conductivity ($\kappa=\kappa_e+\kappa_l$) or at least one of them [7,8]. However, it is usually difficult to greatly enhance the *ZT* value because these transport coefficients are coupled with each other [9-12]. For example, the Seebeck coefficient decreases with increasing carrier concentration while the electrical conductivity is inversed. It is therefore urgently desirable to find new effective approaches to improve *PF* and/or reduce $\kappa$. For example, it was suggested that using low-dimensional systems or nanostructures could be an appropriate way to enhance the thermoelectric performance due to the quantum confinement effect and phonon-boundary scattering [13,14].

---

[*] Author to whom correspondence should be addressed. Electronic mail: phlhj@whu.edu.cn

Since the discovery of graphene in 2004 [15], many other two-dimensional (2D) materials have been intensively studied owing to their unique physical properties and promising application potentials. Recently, a series of superlattice TE materials (AX, A=Si, Ge, Sn, Pb and X=Se, Te) with a unique stacking order of X-A-A-X have been suggested to exhibit remarkable thermoelectric performance [16-18]. Such systems are formed by layered unit (AX)$_2$ through weak van der Waals (vdW) interactions, which makes it possible to exfoliate 2D layers from their bulk materials. Indeed, Sa *et al*. [19] have demonstrated the dynamical stability of such layered unit via first-principles calculations. Moreover, the 2D (PbTe)$_2$ layer [20] was reported to be a good TE material with a maximum *ZT* of 3.0 at 1000 K. Considering the fact that Te element is quite rare on earth, as an alternative, we focus on the PbSe layer in this work and investigate its electronic, phonon and thermoelectric transport properties by using first-principles calculations and Boltzmann theory. It is found that the 2D PbSe exhibits a maximum *p*-type *ZT* of 2.2 at 900 K. Interestingly, we find certain strong coupling between Pb-Pb layers in addition to the weak vdW interactions, which could be used to fine tune the thermoelectric performance of such layered structure.

## 2. Computational method

Within the framework of density functional theory (DFT) [21,22], the electronic properties of the PbSe layer are calculated using the projector augmented wave (PAW) [23] method, as implemented in the Vienna *ab-initio* simulation package (VASP) [24]. The exchange-correlation energy is described by the Perdew-Burke-Ernzerhof (PBE) [25] functional under the generalized gradient approximation (GGA) [26,27]. The energy cutoff is set to 500 eV and the Brillouin zone is sampled with a 15×15×1 Monkhorst-Pack *k*-mesh. For our 2D system, we adopt a vacuum distance of 18 Å to eliminate the interaction between the PbSe layer and its periodic images. To account for the vdW interactions between the Pb-Pb layers, we use the Tkatchenko-Scheffler method with iterative Hirshfeld partitioning (DFT-TS/HI) [28]. The energy convergence criteria is $10^{-6}$ eV for the structure relaxations. Based on the calculated energy band structure, the semiclassical Boltzmann transport theory [29] is applied to evaluate the electronic transport properties including the Seebeck coefficient $S$, the electrical conductivity $\sigma$, and the electronic thermal conductivity $\kappa_e$. In addition, we adopt the deformation potential (DP) theory [30] to evaluate the relaxation time $\tau$, and the effect of spin-orbit coupling (SOC) is explicitly considered in all the calculations.

The phonon dispersion relations of the PbSe layer are obtained by using the harmonic approximation via finite displacement method [31], which is coded in the so-called PHONOPY package [32]. We employ a 7×7×1 and a 4×4×1 supercell in the calculations of the second- and third-order interatomic force constants (IFCs), respectively. The ShengBTE package [33], in which the Peierls-Boltzmann transport equation is implemented, is used for the calculations of phonon transport properties. Furthermore, we adopt a fine *q*-grid mesh of 90×90×1 to guarantee the convergence of lattice thermal conductivity.

## 3. Results and discussion

The top- and side- views of the geometrical structure of the PbSe layer are shown in Figure 1. It forms stacks in the sequence of Se-Pb-Pb-Se with the space group of $P\bar{3}m1$. Similar to graphene, the 2D PbSe exhibits a hexagonal honeycomb lattice. However, there are two buckled sub-layers which may cause intrinsically lower lattice thermal conductivity compared with that of planar 2D materials [34,35]. In addition, we find threefold and sixfold rotational symmetries along the *z*-direction for the Pb and Se sites, respectively. Such a structural characteristic could result in a special band shape and thus affect the electronic transport properties [36]. In order to investigate the effect of vdW interactions between neighboring Pb layers, the so-called TS/HI functional is included in our calculations. Table I lists the optimized lattice parameters of the PbSe layer, where the results without vdW functional are also given for comparison. We see that the lattice constant and the Pb-Se bond length change less by the consideration of vdW interactions. However, there are obvious reductions of the interlayer distance and the Pb-Pb bond length, which confirms the existence of weak vdW forces between neighboring layers in the 2D group-IV chalcogenides. It should be mentioned that the interlayer binding energy of the system is found to be 93 meV/atom, which is significantly larger than that of typical vdW materials such as graphite (31 meV/atom) [37]. On the other hand, the Pb-Pb bond was previously reported to be the strongest in all the 2D group-IV chalcogenides [19]. It is thus reasonable to expect that, in addition to the weak vdW interactions, there should be also certain strong coupling between two sub-layers of the 2D PbSe, as previously found for the SnSe with larger interlayer binding energy (146 meV/atom) [38].

To have a better understanding of such kind of coupling, we plot in Figure 2(a) the differential charge density of the PbSe layer, which is defined as:

$$\Delta\rho_1 = \rho_{\text{total}} - \sum \rho_{\text{atoms}} \qquad (1)$$

where $\rho_{\text{total}}$ is the total charge density of the PbSe layer, and the 2$^{\text{nd}}$ term means superposition of atomic charge densities. In combination with the Bader charge analysis, we find that ~ 0.55 charge is transferred from Pb to Se atoms which suggests that the Pb-Se bond is ionic-like. Besides, there are smaller amount of charges gathered around the Pb atoms, as if they are attracted by neighboring Pb atoms of the other sub-layer. To further investigate the charge distribution between the sub-layers, we plot in Figure 2(b) the interlayer differential charge density given by:

$$\Delta\rho_2 = \rho_{\text{total}} - \sum \rho_{\text{sub-layer}} \qquad (2)$$

where $\rho_{\text{sub-layer}}$ is the charge density of each sub-layer. We see obvious charge accumulation between the Pb-Pb layers, which is reminiscent of covalent-like bond. Such an observation further confirms the existence of strong interlayer coupling in the 2D PbSe system, which is very similar to the layered SnSe exhibiting rather giant anharmonicity [39].

Figure 3(a) plots the energy band structures of the 2D PbSe layer at equilibrium interlayer distances of 2.876 Å. For comparison, the result at a larger distance of 2.930 Å is also shown. We see that both cases exhibit almost the same band structure with a moderate band gap of 0.55 eV. The conduction band minimum (CBM) is located at the $\mathbf{\Gamma}$ point while the valence band maximum (VBM) appears between the $\mathbf{\Gamma}$ and $\mathbf{K}$ points. Besides, there is a valence band extremum (VBE) along the $\mathbf{\Gamma M}$ direction and its energy is very close to that of VBM. Furthermore, we see from Figure 3(b) that both the VBM and VBE show a band degeneracy of three ($N_v = 3$) in the whole Brillouin zone, which is originated from the threefold rotational symmetry discussed above. As a consequence, the PbSe layer can exhibit larger total density of state (DOS) effective mass ($m^*_{total}$) around the VBM, which is calculated by:

$$m^*_{total} = N_v m_{VBM} + N_v m_{VBE} \tag{3}$$

where $m_{VBM}$ and $m_{VBE}$ are the DOS effective mass for the VBM and VBE, respectively. Such kind of multi-valley structure could simultaneously enhance the Seebeck coefficients and the electrical conductivity, as found in previous works [40]. In Figure 3(c), we plot the room temperature Seebeck coefficient (absolutely value) and electrical conductivity as a function of carrier concentration for our 2D PbSe. We first focus on the intrinsic system with interlayer distance of 2.876 Å. In the whole concentration range of $1 \times 10^{18} \sim 10^{21}$ cm$^{-3}$, we see that the $p$-type Seebeck coefficient is obviously larger than that of $n$-type, which is consistent with the multi-valley band structure. Note that at much smaller concentration of $4.45 \times 10^{16}$ cm$^{-3}$, a maximum $p$-type $S$ of 1010 μV/K can be achieved, which is in good agreement with previous theoretical result [41]. On the contrary, we find that the $n$-type system exhibits larger electrical conductivity than that of $p$-type, which is caused by the stronger band dispersion and smaller DOS effective mass of the CBM. Collectively, such findings lead to a higher *p*-type PF of 6.12×10⁻³ W/mK² at optimized concentration of 8.21×10¹⁹ cm⁻³ (Figure 3(d)), where the corresponding Seebeck coefficient can reach 388 µV/K which is larger than those found in many good TE materials such as $Bi_2Te_3$ (320 µV/K) [42] and $Cu_2Se$ (160 µV/K) [43]. Meanwhile, the corresponding electrical conductivity is found to be 4.1×10⁴ S/m, which is also comparable with that of $Bi_2Se_3$ (5.0×10⁴ S/m) [44]. On the other hand, we see from Figure 3(c) and 3(d) that increasing the interlayer distance only leads to small changes of the electronic transport coefficients. For example, the maximum PF is slightly decreased to 5.75×10⁻³ W/mK² at concentration of 8.21×10¹⁹ cm⁻³, where the corresponding Seebeck coefficient and electrical conductivity are 345 µV/K and 4.8×10⁴ S/m, respectively. Such a difference can be attributed to the fact that the system with larger interlayer distance exhibits relatively lower relaxation time ($\tau$), which can be simply obtained by using the DP theory assuming the single parabolic band (SPB) model [45]. For 2D system,

$$\tau = \frac{2\hbar^3 C}{3k_B T m_{dos}^* E^2} \qquad (4)$$

where $C$, $m_{dos}^*$, $T$ and $E$ are the elastic modulus, the averaged DOS effective mass, the absolute temperature, and the deformation potential constant, respectively. Considering the degeneracy of the VBE and the VBM, the averaged DOS effective mass of a single valley is given by $m_{dos}^* = m_{total}^* / N_v$. The calculated room temperature relaxation time and related parameters of PbSe layer are summarized in Table II. Indeed, we see that the intrinsic system with *d* = 2.876 Å exhibits an *n*-type (*p*-type) relaxation time of 80.9 fs (40.6 fs), as compared with that of 68.7 fs (34.5 fs) for the system with *d* = 2.930 Å. The major reason is that the latter exhibits smaller elastic modulus, which is in good agreement with the reduced coupling at increased interlayer distance.

Figure 4(a) plots the phonon dispersion relations of the PbSe layer. Similar to that done for the energy band structure, we consider two different interlayer distances of 2.876 Å and 2.930 Å. We see there is no imaginary frequency for both cases which substantiates their dynamical stability. As 4 atoms are contained in the primitive cell, there are totally 12 phonon branches among which 3 acoustic ones mixed with 3 low-frequency optical ones. Such a kind of hybridization are usually found in many systems with instinctually low lattice thermal conductivity [46]. Additionally, in the high-frequency region, there are 6 flat optical branches which have very small phonon group velocity ($v_{ph}$). Moreover, we find a smaller cutoff frequency of 5.34 THz for the intrinsic PbSe which is comparable with that of Bi$_2$Te$_3$ (4.50 THz) [47]. All these observations indicate that the PbSe layer should possess an ultralow lattice thermal conductivity. On the other hand, if the interlayer distance is increased to 2.930 Å, we find there is almost no change of the 6 high-frequency branches, but an obvious phonon softening of the 6 low-frequency ones which could lead to lower phonon group velocity and larger scattering rate [48]. The physical reason is that the low-frequency branches are dominated by the Pb atoms which govern the interlay interactions, as can be seen from the corresponding phonon density of state (PDOS). Figure 4(b) shows that the lattice thermal conductivity of PbSe layer decreases with increasing temperature and roughly follows a $1/T$ law. In particular, the room temperature lattice thermal conductivity is as small as 2.06 W/mK for the intrinsic PbSe, which is comparable with that of state-of-the-art TE material SnSe (2.60 w/mK) [49]. Such a lower thermal conductivity can be further reduced to 1.59 W/mK when the interlayer distance is increased to 2.930 Å. To have a deep understanding, we plot in Figure 4(c) the atomic displacement parameters (ADP) as a function of temperature. Here we focus on the determinant Pb atoms which exhibit obviously bigger ADP at larger interlayer distance. As a consequence, a relatively lower thermal conductivity can be expected. To go further, we show in Figure 4(d) the total energy of our 2D PbSe at a series of interlayer distance ($d$). It is interesting to find that the energy variation can be well fitted by a polynomial at smaller $d$:

$$E(d) = \sum_{i=0}^{6} a_i (\Delta d)^i \qquad (5)$$

and a Lennard-Jones potential at larger $d$:

$$E(d) = -\frac{A}{d^6} + \frac{B}{d^{12}} \qquad (6)$$

Here $a_i$, $A$ and $B$ are the fitting parameters, and $\Delta d = d - 2.876$ represents the displacement to the equilibrium distance. Such a finding further confirms the fact that there exist strong coupling between the adjacent Pb layers in addition to the weak vdW interactions. It should be noted that the harmonic term in Eq. (5) plays a major role since it has the largest fitting parameter ($a_2$ = 0.162 eV/Å$^2$). Consequently, increasing the interlayer distance would enhance the anharmonicity and in turns reduce the lattice thermal conductivity.

Due to the higher power factor and lower lattice thermal conductivity discussed above, it is reasonable to expect that the PbSe layer could have very favorable TE performance. Indeed, we see from Figure 5 that at equilibrium interlayer distance of 2.876 Å, the PbSe layer exhibits better *p*-type thermoelectric performance than that of *n*-type one, where we find a room temperature *ZT* value of 0.8 and 0.3, respectively. The maximum *ZT* of 2.2 is achieved at temperature of 900 K, with a *p*-type carrier concentration of $7.66 \times 10^{19}$. Such a *ZT* value is obviously higher than those of traditional good TE materials such as PbTe [50,51] and Bi$_2$Te$_3$ [52]. If the interlayer distance is increased to 2.930 Å, the *ZT* value can be further increased to 2.5 at carrier concentration of $1.08 \times 10^{20}$, which is comparable to that of SnSe [53,54]. As discussed above, such enhanced TE performance can be attributed to the fact that reduced interlayer coupling obviously decreases the lattice thermal conductivity (Figure. 4(b)) without much change to the electronic transport properties (Figure 3. (d)).

## 4. Summary

In summary, we demonstrate by first-principles calculations and Boltzmann transport theory that the 2D PbSe could exhibit rather higher *ZT* value, which is caused by the multi-valley band structure around the VBM and unique atomic stacking with buckled sub-layers. Furthermore, we find that weak vdW interactions and strong interlayer coupling coexist between neighboring Pb layers, which could be fine tuned to further enhance the TE performance of the system. In principle, our works provide a new perspective for the application of 2D group-IV chalcogenides as high performance TE materials, and could be generalized to other similar layered systems and 2D hetrostructures.

**Acknowledgements**

We thank financial support from the National Natural Science Foundation of China (Grant Nos. 51772220 and 11574236). The numerical calculations in this work have been done on the platform in the Supercomputing Center of Wuhan University.

**Table I.** The calculated lattice parameters (in unit of Å) of PbSe layer, which include the lattice constants ($a = b$), the bond lengths of Pb-Se ($l_{Pb-Se}$) and Pb-Pb ($l_{Pb-Pb}$), and the interlayer distance ($d$).

| vdW functional | $a = b$ | $l_{Pb-Se}$ | $l_{Pb-Pb}$ | $d$ |
|---|---|---|---|---|
| TS/HI | 4.182 | 2.831 | 3.755 | 2.876 |
| without vdW | 4.176 | 2.834 | 3.812 | 2.950 |

**Table II.** The room temperature relaxation time and related parameters of PbSe layer at two different interlayer distances.

| Interlayer distance (Å) | Carrier type | $C(eV/Å^2)$ | $m^*_{dos}/m_e$ | $E(eV)$ | $\tau(fs)$ |
|---|---|---|---|---|---|
| 2.876 | Electron | 4.276 | 0.150 | −6.76 | 80.9 |
|  | Hole | 4.276 | 0.664 | −4.79 | 40.6 |
| 2.930 | Electron | 3.768 | 0.153 | −6.79 | 68.7 |
|  | Hole | 3.768 | 0.664 | −4.68 | 34.5 |

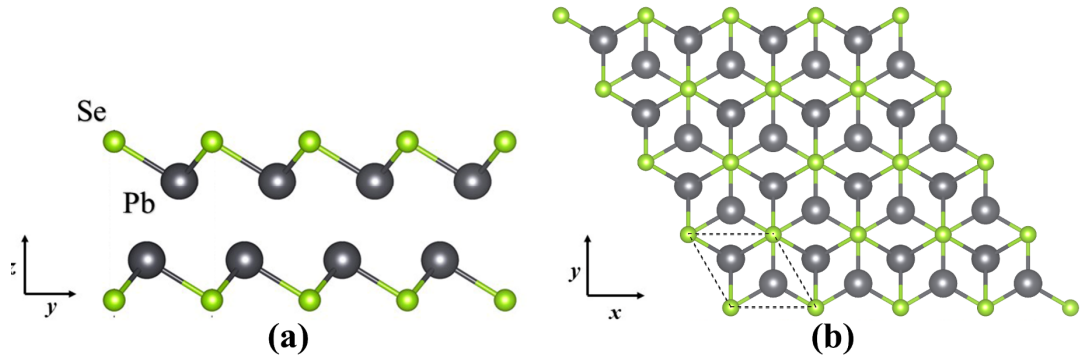

**Figure 1**. The crystal structure of PbSe layer (a) side-view and (b) top-view. The dashed lines denotes the primitive cell.

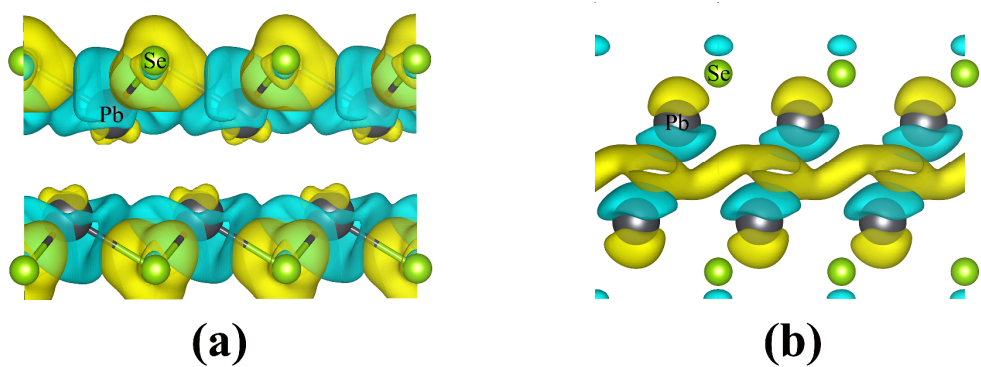

**Figure 2.** The differential charge density defined by (a) Eq. (1) and (b) Eq. (2), where the yellow and the cyan denote obtaining and losing charges, respectively.

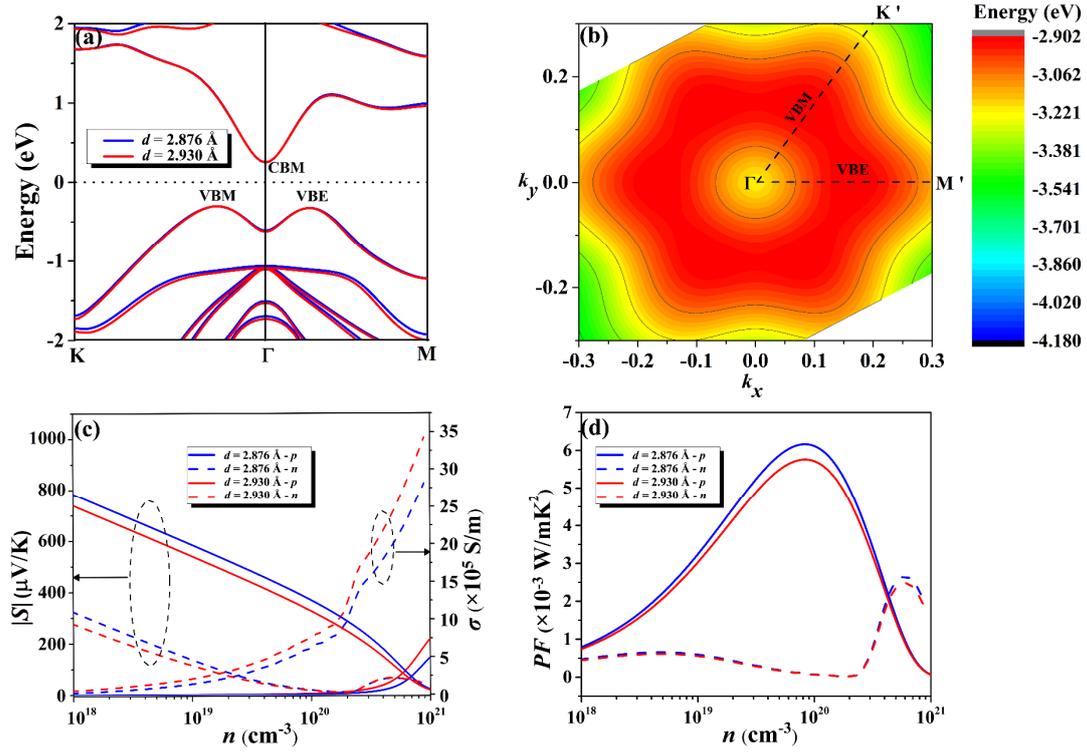

**Figure 3.** (a) The band structure of PbSe layer at two different interlayer distances. (b) The energy distribution of top valence bands in the whole Brillouin zone. (c) The room temperature electrical conductivity and Seebeck coefficient as a function of carrier concentration along the *x*-direction. (d) The room temperature power factor as a function of carrier concentration along the *x*-direction.

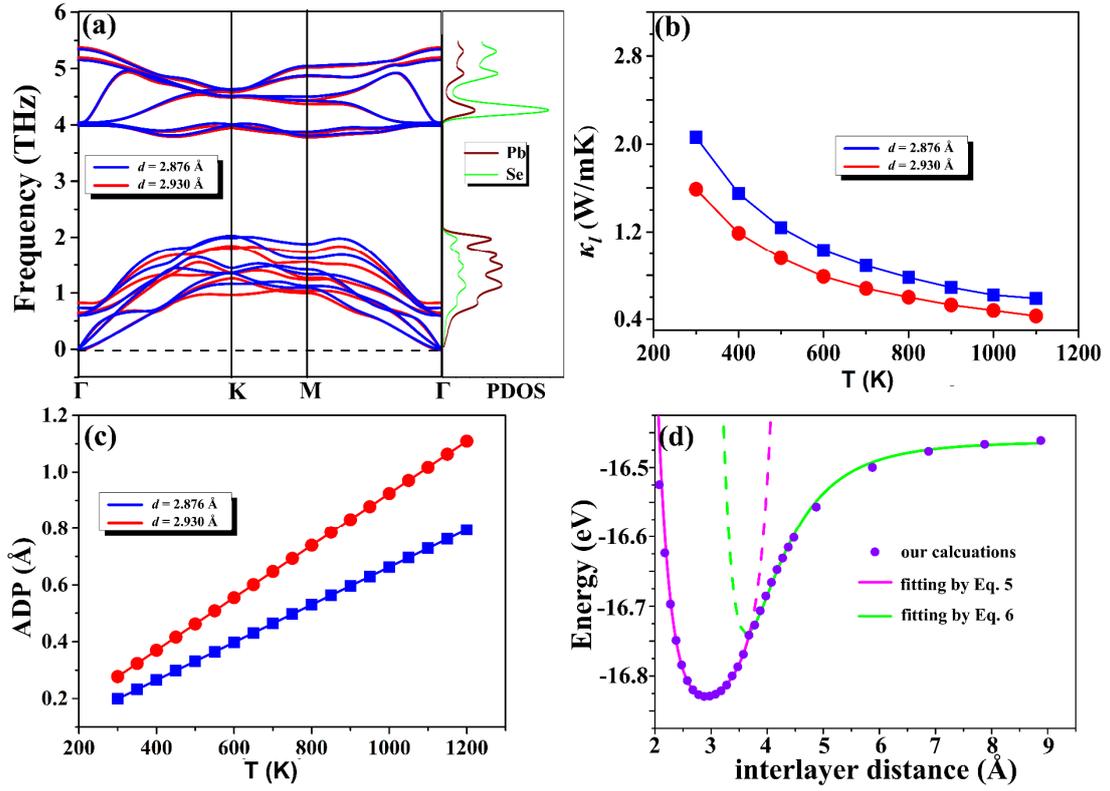

**Figure 4.** (a) The phonon dispersion relations of PbSe layer at two different interlayer distances, where the phonon DOS is also given for intristic system. (b) The temperature dependent lattice thermal conductivity. (c) The atomic displacement parameters of Pb atoms as a function of temperature. (d) The energy variation as a function of interlayer distance.

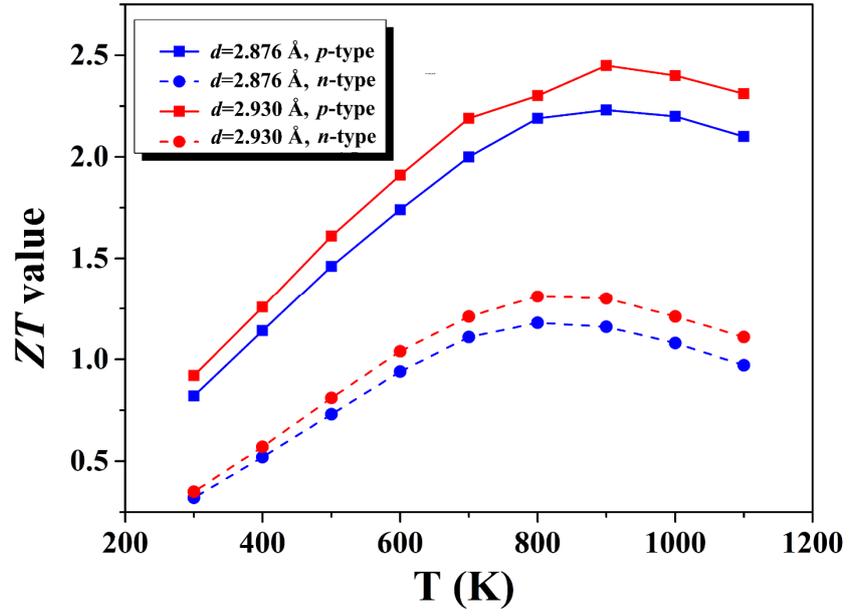

**Figure 5.** The temperature dependence of the *ZT* values for the PbSe layer at two different interlayer distances.

**Reference**

[1] C. B. Vining, *Nat. Mater.* **8**, 83 (2009).

[2] A. Majumdar, *Science* **303**, 777 (2004).

[3] H. J. Goldsmid, *Introduction to Thermoelectricity*, (Springer, New York, 2009).

[4] Y. Lu, T. Sun and D. B. Zhang, *Phys. Rev.* B **97**, 174304 (2018).

[5] W. D. Liu, Z. G. Chen and J. Zhou, *Adv. Energy Mater.* **8**, 1800056 (2018).

[6] Z. G. Chen, X. Shi, L. D. Zhao and Z. Jin, *Prog. Mater. Sci.* **97**, 283 (2018).

[7] G. J. Snyder and E. S. Toberer, *Nat. Mater.* **7**, 105 (2008).

[8] G. Schierning, R. Chavezl and R. Schmechell, *Transl. Mater. Res.* **2**, 025001 (2015).

[9] G. Slack, in: D. M. Rowe (Ed.), *CRC Handbook of Thermoelectrics*, CRC Press, Boca Raton, FL, (1995).

[10] J. R. Sootsman, D. Y. Chung and M. G. Kanatzidis, *Angew. Chem. Int. Ed.* **48**, 8616 (2009).

[11] C. J. Vineis, A. Shakouri, A. Majumdar and M. G. Kanatzidis, *Adv. Mater.* **22**, 3970 (2010).

[12] M. G. Kanatzidis, *Chem. Mater.* **22**, 648 (2018).

[13] L. D. Hicks and M. S. Dresselhaus, *Phys. Rev.* B **47**, 12727 (1993).

[14] L. D. Hicks and M. S. Dresselhaus, *Phys. Rev.* B **47**, 16631 (1993).

[15] K. S. Novoselov, A. K. Geim, S. Morozov, D. Jiang, Y. Zhang, S. A. Dubonos, I. Grigorieva and A. Firsov, *Science*, **306**, 666 (2004).

[16] B. Sa, J. Zhou, Z. Sun, J. Tominaga and R. Ahuja, *Phys. Rev. Lett.* **109**, 096802 (2012).

[17] J. Tominaga, A. Kolobov, P. Fons, T. Nakano and S. Murakami, *Adv. Mater. Interfaces* **1**, 1300027 (2014).

[18] Y. Saito, J. Tominaga, P. Fons, A. Kolobov and T. Nakano, *Phys. Status Solidi RRL* **8**, 302 (2014).

[19] B. S. Sa, Z. M. Sun and B. Wu, *Nanoscale* **8**, 1169 (2016).

[20] C. Y. Sheng, D. D. Fan and H. J. Liu. *Phys. Lett.* A **384**, 126044 (2020).

[21] G. Kresse and J. Hafner, *Phys. Rev.* B **47**, 558 (1993).

[22] G. Kresse and J. Hafner, *Phys. Rev.* B **49**, 14251 (1994).

[23] J. Hafner, *J. Comput. Chem.* **29**, 2044 (2008).

[24] G. Kresse and J. Furthmüller, *Phys. Rev.* B **54**, 11169 (1996).

[25] G. Kresse and D. Joubert, *Phys. Rev.* B **59**, 1758 (1999).

[26] P. E. Blöchl, O. Jepsen and O. K. Andersen, *Phys. Rev.* B **49**, 16223 (1994).

[27] P. E. Blöchl, *Phys. Rev.* B **50**, 17953 (1994).

[28] A. Tkatchenko and M. Scheffler, *Phys. Rev. Lett.* **102**, 073005 (2009).

[29] G. K. H. Madsen and D. J. Singh, *Comput. Phys. Commun.* **175**, 67 (2006).

[30] J. Bardeen and W. Shockley, *Phys. Rev.* **80**, 72 (1950).

[31] S. Baroni, S. De Gironcoli, A. Dal Corso and P. Giannozzi, *Rev. Mod. Phys.* **73**, 515 (2001).

[32] A. Togo, F. Oba and I. Tanaka, *Phys. Rev.* B **78**, 134106 (2008).

[33] W. Li, J. Carrete, N. A. Katcho and N. Mingo, *Comput. Phys. Commun.* **185**, 1747 (2014).

[34] X. Gu. and R. G. Yang, *Appl. Phys. Letts.* **105**, 131903 (2014).

[35] X. Gu. and R. G. Yang, *Appl. Phys. Letts.* **117,** 025102 (2015).

[36] D. D. Fan, H. J. Liu, L. Cheng, J. Zhang, P. H. Jiang, J. Wei, J. H. Liang and J. Shi, *Phys. Chem. Chem. Phys.* **19**, 12913 (2017).

[37] Z. Liu, J. Z. Liu, Y. Cheng, Z. H. Li, L. Wang, Q. S. Zheng, *Phys. Rev.* B **85**, 205418 (2012).

[38] H. Y. Song, and J. T. Lv, *Chem. Phys. Lett.* **695,** 200 (2018).

[39] C. W. Li, J. Hong and A. F. May, *Nat. Phys.* **11**, 1063 (2015).

[40] Y. Z. Pei, X. Y. Shi, A. Lalonde, H. Wang, L. D. Chen and G. J. Snyder, *Nature* **473**, 66 (2011).

[41] X. L. Zhu, C. H. Hou, P. Zhang, P. F. Liu, G. F. Xie and B. T. Wang, *J. Phys. Chem.* C **124**, 1812 (2020).

[42] H. W. Jeon, H. P. Ha, D. B. Hyun and J. D. Shim, *J. Phys. Chem. Solids* **52**, 579 (1991).

[43] H. L. Liu, X. Yuan, P. Lu, X. Shi, F. F. Xu, Y. He, Y. S. Tang, S. Q. Bai, W. Q. Zhang, L. D. Chen, Y. Lin, L. Shi, H. Lin, X. Y. Gao, X. M. Zhang, H. Chi and C. Uher, *Adv. Mater.* **25**, 6007 (2013).

[44] L. Cheng, H. J. Liu, J. Zhang, J. Wei, J. H. Liang, J. Shi and X. F. Tang, *Phys. Rev.* B **90**, 085118 (2014).

[45] D. M. Rowe and C. M. Bhandari, *Modern Thermoelectric*, Reston Publishing Company, Inc., Reston Virginia, p. **26**.

[46] D. Wee, B. Kozinsky, N. Marzari and M. Fornari, *Phys. Rev.* B **81**, 045204 (2010).

[47] B. Qiu, X. L. Ruan, *Phys. Rev.* B **80**, 165203 (2009).

[48] C. J. Zhou, Y. Yu, Y. K. Lee, O. Cojocaru-Miredin, B. Yoo, S. P. Cho, J. Im, M. Wutting, T. Hyeon, and I. Chung, *J. Am. Chem. Soc.* **140**, 15535 (2018).

[49] A. Shafique and Y. H. Shin, *Sci. Rep.* **7**, 506 (2017).

[50] H. J. Wu, L. D. Zhao, F. S. Zheng, D. Wu, Y. L. Pei, X. Tong, M. G. Kanatzidis, and J. Q. He, *Nat. Commun.* **5**, 4515 (2014).

[51] Z. W. Chen, Z. Z. Jian, W. Li, Y. J. Chang, B. H. Ge, R. Hanus, J. Yang, Y. Chen, M. X. Huang, G. J. Snyder and Y. Z. Pei, *Adv. Mater.* **29**, 1606768 (2017).

[52] B. Poudel, Q. Hao, Y. Ma, Y. C. Lan, A. Minnich, B. Yu, X. Yan, D. Z. Wang, A. Muto, D. Vashaee, X. Y. Chen, J. M. Liu, M. S. Dresselhaus, G. Chen and Z. F. Ren, *Science* **320**, 634 (2008).

[53] L. D. Zhao, S. H. Lo, Y. S. Zhang, H. Sun, G. J. Tan, C. Uher, C. Wolverton, V. P. Dravid and M. G. Kanatzidis, *Nature* **508**, 373 (2014).

[54] L. D. Zhao, G. J. Tan, S. Q. Hao, J. Q. He, Y. L. Pei, H. Chi, H. Wang, S. K. Gong, H. B. Xu, V. P. Dravid, C. Uher, G. J. Snyder, C. Wolverton and M. G. Kanatzidis, *Science* **351**, 141 (2016).